\documentclass[aps,pre,twocolumn,float]{revtex4}
\usepackage{amsmath,bm,epsfig}

\def\Fbox#1{\vskip1ex\hbox to 8.5cm{\hfil\fboxsep0.3cm\fbox{%
  \parbox{8.0cm}{#1}}\hfil}\vskip1ex\noindent}  

\newcommand{\B}[1]{{\bm{#1}}}
\newcommand{\C}[1]{{\mathcal{#1}}}    
\newcommand{\sFrac}[2]{{\textstyle\frac{#1}{#2}}}
\let \= \equiv \let\*\cdot \let\~\widetilde \let\^\widehat \let\-\overline

\begin{document}
\title{Plasticity-Induced Anisotropy in Amorphous Solids: the Bauschinger Effect}
\author{Smarajit Karmakar, Edan Lerner and Itamar Procaccia}
\affiliation{Department of Chemical Physics, The Weizmann
Institute of Science, Rehovot 76100, Israel}
\begin{abstract}
Amorphous solids that underwent a strain in one direction such that they responded in a plastic manner `remember'
that direction also when relaxed back to a state with zero mean stress. We address the question `what is the order
parameter that is responsible for this memory?' and is therefore the reason for the different subsequent responses
of the material to strains in different directions. We identify such an order parameter which is readily measurable, we
discuss its trajectory along the stress-strain curve, and propose that it and its probability distribution function
must form a necessary component of a theory of elasto-plasticity.
\end{abstract}
\maketitle
\section {Introduction}
\label{intro}
\begin{figure}
 \centering
\includegraphics[scale = 0.50]{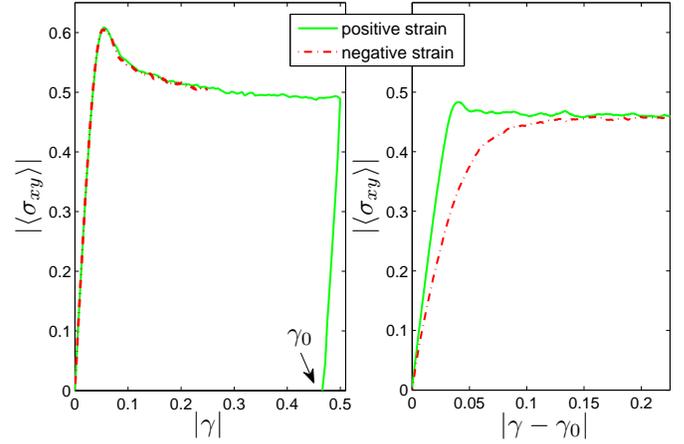}
\caption{Color online: Stress-strain curves. Left panel: starting the experiment from a freshly prepared sample results in a symmetric trajectory for $\gamma\to -\gamma$. Right panel: starting the experiment from the zero-stress state with $\gamma=\gamma_0$ results in an asymmetric trajectory, see text for details. Data was averaged over 500 independent stress-strain curves at $T=0.01$ where temperature is measured in units of $\epsilon/k_B$, see Subsect \ref{model} below.}
\label{stress-strain}
\end{figure}

An amorphous solid which is freshly produced by cooling a glass-forming system from high
to low temperature is isotropic up to small statistical fluctuations. Put under an external strain, its stress vs. strain curve should exhibit symmetry for positive or negative strains. This is not the case for the same amorphous solid after it had been already strained such that its stress exceeded its yield-stress where plastic deformations become numerous, resulting in an elasto-plastic flow state. The phenomenon is clearly exhibited in Fig.~\ref{stress-strain}. A typical averaged stress-strain curve for a 2-dimensional model
amorphous solid (see below for numerical details) starting from an ensemble of freshly prepared homogenous states is shown in the left panel, with a symmetric trajectory for positive or negative shear strain. Once in the steady flow state, each system in the ensemble is brought back to a zero-stress state, which serves as the starting point for a second experiment in which a positive and negative strain is put on the system as shown
in the right panel of Fig.~\ref{stress-strain}. Even though the initial ensemble is prepared to have zero mean stress, the
average trajectory now is asymmetric, with positive strain exhibiting `strain hardening' \cite{foot1}, but reaching the same level of steady state flow-stress, whereas, the
negative strain results in a `strain softening' and a faster yield with eventually reaching the same value of steady-state flow-stress (in absolute value). This simple phenomenon, sometime referred to as the Bauschinger effect \cite{90BCH},
shows that the starting point $\gamma_0$ for the second experiment (referred below as the Bauschinger point) retains a memory of the loading history, some form of anisotropy,
which is the subject of this article. We stress that the issue under study is different from anisotropic elasticity which
stems from, say, a lattice anisotropy of a crystalline solid. Here the systems under study are amorphous, and nevertheless
develop a strain induced anisotropy which is much more subtle to identify and quantify.

How to identify the order parameter which is responsible for the anisotropy underlying the Bauschinger effect is a
question that hovers in the elasto-plastic community for some while \cite{79AK,79Arg,82AS}. One obvious concept, i.e. of `back stress' \cite{98Sur} or `remnant stress' for explaining the asymmetry seen in the second experiment in Fig. \ref{stress-strain} can be ruled out simply by verifying that the initial point has zero mean stress. A more sophisticated proposition is embodied in the `shear-transformation zone' theory (STZ) in which it is conjectured that plasticity occurs in localized regions whose densities differ for positive and negative strains, denoted $n_+$ and $n_-$ \cite{98FL,07BLP}. The normalized difference between these, denoted as $m$, is a function of the loading history and can, in principle, characterize the anisotropy that we are seeking. Unfortunately the precise nature of the STZ's was never clarified, and it is unknown how to measure either
$n_+$, $n_-$ or $m$, making it quite impossible to put this proposition under a direct test. More recently it was proposed
that the sought after anisotropy can be characterized in granular matter by the fabric tensor $\B F =\langle \B n \B n\rangle$ which captures the mean orientation of the contact normals, $\B n$, through the spatial average of their diadic product \cite{09RVTBR}. This order parameter was generalized for silica glass where $\B n$ was chosen as a unit vector in the direction of the vector distance between Si atoms, disregarding the oxygens. Attempting to test this proposition in the context of the best-studied model of glass-forming, i.e. a binary mixtures of point particles with two interaction lengths, or in the case of multi-dispersed point particles (see below for details), did not reveal any systematic signature of anisotropy. We thus conclude that this order parameter is not sufficiently general to be of universal use in the development of the theory of elasto-plasticity, and that the question of identifying a missing order parameter remains open.
\begin{figure}[h]
 \centering
\includegraphics[scale = 0.5]{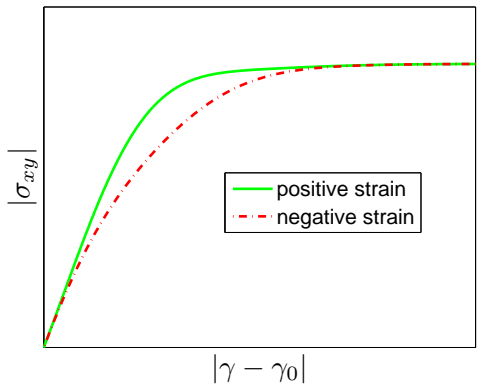}
\caption{Color Online: Model stress-strain curves as obtained from $\sigma\sim  \sigma_\infty\tanh(\mu|\gamma-\gamma_0|/\sigma_\infty) + \beta |\gamma-\gamma_0|^2e^{-|\gamma-\gamma_0|^2}$ with $\beta$ positive.}
\label{toymodel}
\end{figure}

\section {The proposed order parameter}
\label{B2}
To see what may serve as a general order parameter we examine first the situation
with the isotropic amorphous solid which is obtained after a quench without any loading history. Denoting the shear stress
by $\sigma$ and the shear strain by $\gamma$, we observe that isotropy dictates that all the even derivatives
$d^{2n}\sigma/d \gamma^{2n}$ must vanish by symmetry. For example a function that can model the stress-strain curve
with this constraint in mind may be $\sigma \sim \sigma_\infty\tanh(\mu\gamma/\sigma_\infty)$ where $\sigma_\infty$ is the flow stress (the mean value of the stress in the elasto-plastic steady state),
and $\mu$ is the shear modulus. For $\gamma\to -\gamma$ this
function is perfectly anti-symmetric as required for an isotropic system. Imagine now that we add even derivatives
to this function, say $\sigma\sim  \sigma_\infty\tanh(\mu\gamma/\sigma_\infty) + \beta \gamma^2e^{-\gamma^2}$ with $\beta$ having the dimension of stress. The effect will be to change the stress-strain curve as seen in Fig. \ref{toymodel}, which is quite reminiscent of the Bauschinger effect. We therefore propose that it is advantageous to focus on the even derivatives of $\sigma$ vs. $\gamma$, with the most important one being the second derivative. Note nevertheless that the {\em mechanism} leading to the existence of a second derivative in our systems is not obvious in this simple model.
The second derivative appears due to plastic deformations whose effect adds up to breaking the isotropic symmetry of the
freshly quenched state. We will show that at the Bauschinger point the second derivative is non-zero due to existing
closer mechanical instabilities in one straining direction than in the opposite.

\subsection{Statistical Mechanics} Under external loads the displacement field $\B v$ describes how a material point
moved from its equilibrium position. The strain field is defined (to second order) as
\begin{equation}\label{strainTensorDefinition1}
\epsilon_{\alpha\beta} \equiv \frac{1}{2}
\left(
\frac{\partial v_\alpha}{\partial x_\beta} + \frac{\partial v_\beta}{\partial x_\alpha} +
\frac{\partial v_\nu}{\partial x_\alpha}\frac{\partial v_\nu}{\partial x_\beta}
\right)\ ,
\end{equation}
where here and below repeated Greek indices are summed upon. We expand the free energy density ${\cal F}/V$ up to a constant in terms of the strain tensor
\begin{equation}
\frac{ {\cal F} }{V} = C_1^{\alpha\beta}\epsilon_{\alpha\beta} + \sFrac{1}{2}C_2^{\alpha\beta\nu\eta}\epsilon_{\alpha\beta}\epsilon_{\nu\eta}
+ \sFrac{1}{6}C_3^{\alpha\beta\nu\eta\kappa\chi}
\epsilon_{\alpha\beta}\epsilon_{\nu\eta}\epsilon_{\kappa\chi}\ .
\end{equation}
The mean stress is defined as $\sigma_{\alpha\beta}  \equiv \sFrac{1}{V}\sFrac{\partial {\cal F}}{\partial \epsilon_{\alpha\beta}} $, and
\begin{equation}\label{stressExpansion}
 \sigma_{\alpha\beta}   = C_1^{\alpha\beta} +
C_2^{\alpha\beta\nu\eta}\epsilon_{\nu\eta}+
\sFrac{1}{2}C_3^{\alpha\beta\nu\eta\kappa\chi}\epsilon_{\nu\eta}\epsilon_{\kappa\chi}\ .
\end{equation}
In our simulations we apply a simple shear deformation using the transformation of coordinates according to
\begin{equation}\label{affineTransformation}
\begin{split}
x_i & \to x_i + \delta\gamma y_i \ , \\
y_i & \to y_i \ , (z_i\to z_i \mbox{ in 3 dimensions})
\end{split}
\end{equation}
where $\delta\gamma = \gamma - \gamma^*$ is a small strain increment from any reference strain $\gamma^*$.
The explicit 2D strain tensor following Eq.~(\ref{strainTensorDefinition1}) is
\begin{equation}
\epsilon = \frac{1}{2}\left(
\begin{array}{cc}
0&\delta\gamma\\
\delta\gamma&\delta\gamma^2
\end{array}
\right)\ ,
\end{equation}
with an obvious generalization in 3D. Since $\epsilon_{xx}=0$, the mean shear stress reduces to the form (equally
valid in 2D and 3D)
\begin{equation}
\sigma_{xy} = C_1^{xy} + C_2^{xyxy}\delta\gamma
+\sFrac{1}{2}(C_2^{xyyy} +  C_3^{xyxyxy})\delta\gamma^2  + {\cal O}(\delta\gamma^3)\ .
\end{equation}
As discussed above, in isotropic systems where $ \sigma_{xy}$ is antisymmetric in $\delta\gamma$,
$C_1^{xy} = 0$ and the sum $C_2^{xyyy} + C_3^{xyxyxy} = 0$.
Our proposition is to use the athermal limit of this sum as the characterization of the anisotropy that we seek.

\subsection{Models and numerical procedures}
\label{model}
Below we employ a model system with point particles of equal mass $m$ and positions $\B r_i$ in two and three-dimensions, interacting via a pairwise potential of the form
\begin{equation}\label{potential}
\phi\left(\!\frac{r_{ij}}{\lambda_{ij}}\!\right) =
\left\{ \begin{array}{ccl} \!\!\varepsilon\left[\left(\frac{\lambda_{ij}}{r_{ij}}\right)^{k} + \displaystyle{\sum_{\ell=0}^{q}}c_{2\ell} \left(\frac{r_{ij}}{\lambda_{ij}}\right)^{2\ell}\right] &\! , \! & \frac{r_{ij}}{\lambda_{ij}} \le x_c \\ 0 &\! , \! & \frac{r_{ij}}{\lambda_{ij}} > x_c \end{array} \right., \end{equation} where $r_{ij}$ is the distance between particle $i$ and $j$, $\varepsilon$ is the energy scale, and $x_c$ is the dimensionless length for which the potential will vanish continuously up to $q$ derivatives. The coefficients $c_{2\ell}$ are given by \begin{equation} c_{2\ell} = \frac{(-1)^{\ell+1}}{(2q-2\ell)!!(2\ell)!!}\frac{(k+2q)!!}{(k-2)!!(k+2\ell)}x_c^{-(k+2\ell)}.
\end{equation}
We chose the parameters $k=10$, $q=2$ and $x_c=1.385$. In the three-dimensional simulations each particle $i$ is assigned an interaction parameter $\lambda_i$ from a normal distribution with mean $\langle \lambda \rangle$ and
$\lambda_{ij} = \frac{1}{2}(\lambda_i + \lambda_j)$. The variance is governed by the poly-dispersity parameter $\Delta = 15\%$ where $\Delta^2 = \frac{\langle \left(\lambda_i -
\langle \lambda \rangle \right)^2\rangle}{\langle \lambda \rangle^2} $.
In the two dimensional simulations we use the same potential but choose a binary mixture model with `large' and
`small' particles such that $\lambda_{LL} =1.4$, $\lambda_{LS}=1.18$ and $\lambda_{SS}=1.00$.
Below the units of length are $\lambda=\lambda_{SS}$ in 2D and $\lambda=\langle \lambda\rangle$ in 3D. We measure energy, mass and temperature in units of $\varepsilon$, $m$ and ($\varepsilon/k_B$) respectively. The unit of time is $\tau=\sqrt{m\lambda^2/\varepsilon}$. In the 3D simulations below the mass density $\rho\equiv m N/V=1.3$, whereas in 2D $\rho=0.85$. In all cases the boundary conditions are periodic and thermostating is
achieved with the Berendsen scheme \cite{91AT}. We employ the sllod equations of motion for imposing deformations, and integrate them using a standard leap-frog algorithm \cite{91AT}. The strain rate is chosen to be $\dot{\gamma} = 10^{-4}\tau^{-1}$ for all simulations described below. Initial configurations were prepared by equilibrating at least 1000 independent systems in the supercooled temperature regime, followed by quenching to the target temperature at a rate of
$10^{-4}\frac{\varepsilon}{k_B\tau}$. If not stated otherwise all the simulation below were obtained with systems of
$N=20164$ in 2-dimensions and $N=16384$ in 3-dimensions.

We choose to measure the sum
\begin{equation}
B_2(\gamma^*) \equiv \!\lim_{T\to 0}\! \left[C_2^{xyyy} + C_3^{xyxyxy}\right] \!= \!\lim_{T\to 0}\left. \frac{d^2 \sigma_{xy}}{d \gamma^2}
\right|_{\gamma = \gamma^*},
\label{defB2}
\end{equation}
using an athermal, quasi-static scheme \cite{09LP}. This scheme
consists of imposing the affine transformation (\ref{affineTransformation}) to each particle of a minimized configuration,
followed by another potential energy minimization under Lees-Edwards boundary condition \cite{91AT}.
 In athermal quasi-static conditions ($T\to 0, \quad \dot\gamma\to 0$), the system lives in local minima, and follows strain-induced changes of the potential energy surface~\cite{10KLLP}. Therefore, the particles do not follow homogeneously the macroscopic strain, and their positions change as $\B r_i\to \B r'_i+\B u_i$, where $\B u_i$ denotes non-affine displacements. Around some stable reference state at $\gamma=\gamma^*$, the field $\B u_i$, the potential energy $U$, and internal stress $\sigma_{xy}$ are smooth functions of $\gamma$.
 We choose the stopping criterion for the minimizations to be $|\nabla_iU| < 10^{-9}\frac{\varepsilon}{\lambda}$ for every coordinate $x_i$. Within this method one can obtain purely elastic trajectories of stress vs strain \cite{09LP}. To measure $B_2$ of a given configuration of our molecular dynamics simulation at any temperature $T$, we first cool that configuration to $T = 10^{-3}$ using molecular dynamics during a time interval of 50$\tau$. This
chosen temperature is sufficiently low to exclude any thermal activation on the time scale of the simulations. This initial treatment brings our configuration to an elastically stable state i.e. a minimum of the potential energy landscape. Without doing so one can find oneself in the vicinity of
a saddle point for which the athermal elastic moduli have no clear meaning. We then apply the athermal quasi-static scheme to measure the finite differences approximation to $\frac{d ^2 \sigma_{xy}}{d \gamma^2} \approx \frac{ \sigma_{xy}(\delta\gamma)
 + \sigma_{xy}(-\delta\gamma) - 2\sigma_{xy}(0)}{\delta\gamma^2}$ by sampling a small elastic trajectory, using strain increments of $\delta\gamma = 2.5\times 10^{-6}$. We have checked that stricter stopping criteria for the minimizations or smaller strain increments do not significantly alter our results. We emphasize that although we measure $B_2$ in the athermal limit, the configurations on which we perform this measurement are sampled from various finite temperatures, see below. This athermal measurement is motivated by the requirement to probe the purely mechanical response, excluding thermal activation effects on the measurement from the discussion. Using this method we can compute $B_2$ at any point of the trajectory. Note that $B_2$ is still a strong function of the temperature from which the configuration was taken, and this is because the organization of the particles depends on the temperature. We reiterate that $B_2$ is {\bf not} the second derivative of the averaged stress-strain curve, but rather the mechanical response of the underlying inherent structure which is sampled at a given temperature.

 \begin{figure}
 \centering
\includegraphics[scale = 0.62]{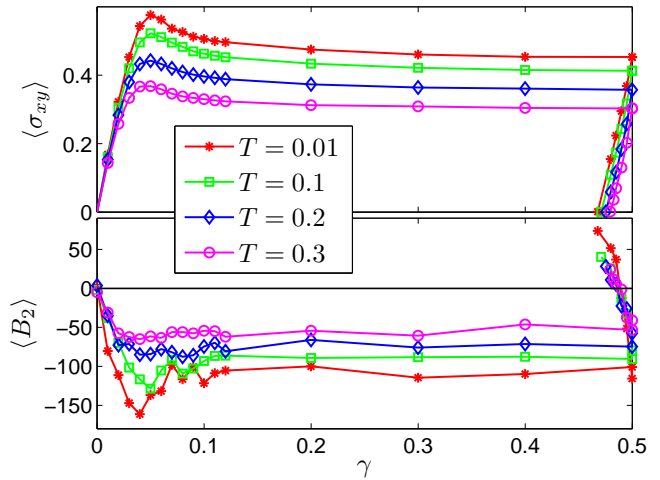}
\caption{Color Online: Upper panel: Trajectories of stress vs. strain for four different temperature at the same strain rate $\gamma=10^{-4}$. Lower panel: the corresponding values of $B_2$ as a function of strain. Data was averaged over 1000 independent stress-strain curves at each temperature. Note that $B_2$ is negative even when the averaged stress-strain curve has a positive curvature, see text for discussion.}
 \label{B2vsgamma}
 \end{figure}

\section {Results and discussion}
\begin{figure}
\centering
\includegraphics[scale = 0.52]{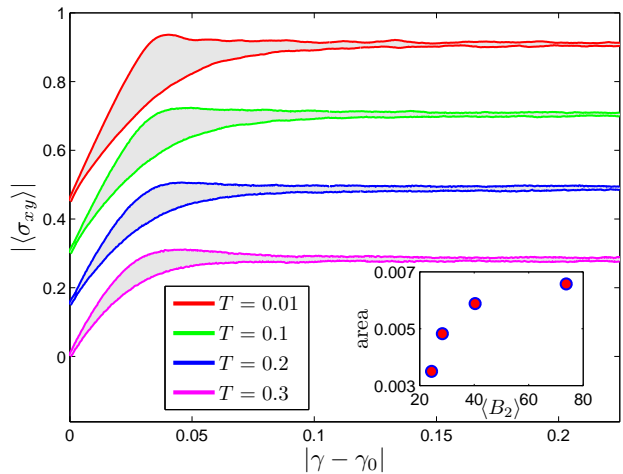}
\caption{Color Online: The Bauschinger effect for the four temperatures shown in Fig. \ref{B2vsgamma}, increasing from top to bottom. The
trajectories are displaced by fixed amount ($\Delta |\sigma_{xy}|=0.15$) for clarity. Note the reduction
of the effect with increasing temperature. Data was averaged over 500 independent stress-strain curves at each temperature. Inset: the shaded area of difference between the stress-strain curves with positive and negative strain as a function of
$B_2$. The magnitude of the Bauschinger effect saturates for $T\to 0$.}
 \label{sigvsgamma}
 \end{figure}

The average trajectories of both stress vs
  strain (upper panel) and $B_2$ vs strain (lower panel) are shown for four different bath temperatures in Fig. \ref{B2vsgamma}; the system was strained until $\gamma=1/2$ and then strain was reversed until the mean
 stress dropped to zero. The strain value was then $\gamma_0$ from which the experiment in Fig. 1 right panel was
 started with positive and negative straining with respect to $\gamma_0$. The resulting trajectories of stress vs. strain are shown in Fig. \ref{sigvsgamma} for the 2D system at the same four values of the temperature as in Fig. \ref{B2vsgamma}.
 We observe that the value of $B_2$ at the point of zero stress $\gamma_0$ reduces when the temperature increases, and
in accordance with that the magnitude of the Bauschinger effect goes down as seen in Fig. \ref{sigvsgamma}.

\begin{figure}
 \centering
\includegraphics[scale = 0.45]{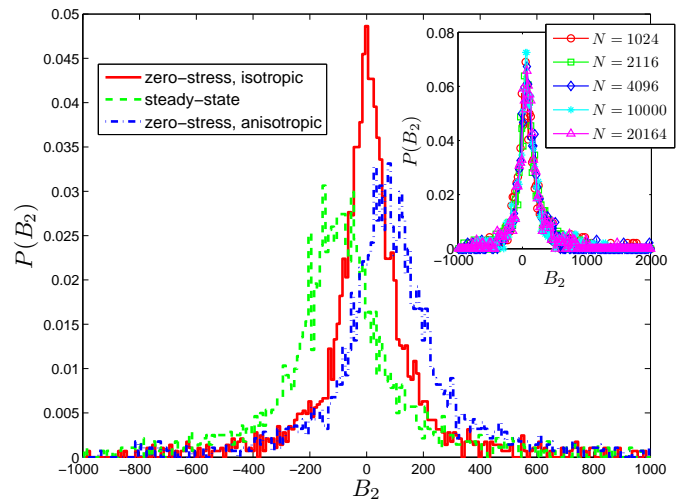}
\caption{Color Online: Probability distribution function of $B_2$ at various points on the stress-strain trajectory.
Data was collected from 3000 independent stress-strain trajectories at $T=0.01$.
In red (continuous) line we draw the symmetric pdf of the freshly prepared samples with $\gamma=0$. In green
(dashed) line we show the pdf in the steady state, where it gains a negative asymmetry. In blue (dashed-dotted)
line we see the pdf at the Bauschinger point $\gamma=\gamma_0$ where it gained a positive asymmetry. The dynamics
of these pdf's and their means are correlated with the shapes of the stress-strain curves and are proposed to be
a crucial ingredient in any theory of elasto-plasticity. Inset: the $N$ dependence of the pdf at the Bauschinger point
$\gamma=\gamma_0$. Data was averaged over 1000 independent samples for each system size.}
 \label{pdf}
 \end{figure}

We can draw the conclusion that the magnitude of $B_2$ is correlated with the amplitude of the Bauschinger effect (measured as the area of difference between the positive and negative stress-strain curves, see inset in Fig. \ref{sigvsgamma}). But even more detailed information which is highly relevant to the elasto-plastic behavior can be gleaned from the {\em probability distributions functions (pdf's)} of $B_2$. These pdf's have rich dynamics along the stress-strain curves, as can be seen in Fig. \ref{pdf}.
When measured in the isotropic zero-stress systems that are freshly quenched the distribution is symmetric as expected,
with zero mean. In the elasto-plastic steady state the distribution moved to have a negative mean, in accordance
with the low panel of Fig.~\ref{B2vsgamma}. In Sect. \ref{theory} we show that this distribution must send a tail towards $-\infty$ to accommodate the sharp changes in first
derivative (the shear modulus) due to the proximity of mechanical instabilities in the form of plastic drops \cite{06ML,10KLLP}.
At the Bauschinger point $\gamma_0$ the mean stress is zero, but the pdf of $B_2$ gains a positive asymmetry, sending
a tail towards $+\infty$, signalling a proximity to a plastic event in the negative straining direction. In the inset of Fig. \ref{pdf} we exhibit the size dependence of the pdf at the Bauschinger point, to show that the asymmetry and the general shape of the pdf is quite independent of the number of particles $N$,
always having long tails,  indicating that near the
Bauschinger point $\gamma_0$ there are close-by lurking plastic instabilities that are heralded by the tail of our pdf.

\begin{figure}[!t]
 \centering
\includegraphics[scale = 0.50]{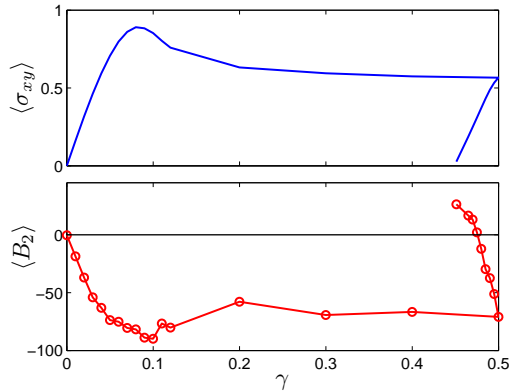}
\caption{A representative averaged stress-strain curve (averaged over 600 independent trajectories) for the 3-dimensional model (upper panel) and the
corresponding trajectory of $B_2$ in the lower panel for $T=0.01$.}
\label{3D}
\end{figure}

To confirm that the qualitative findings reported above remain unchanged in 3-dimensions we repeated
similar simulation for the model described above. In Fig. \ref{3D} we present a representative averaged stress-strained
curve in the upper panel and the corresponding trajectory of $B_2$, both at $T=0.01$.

\section{The divergence of $B_2$ near mechanical instabilities in the form of plastic events}
\label{theory}
To see that $B_2$ must reach $\pm \infty$ when the system goes through a plastic deformation recall that as long as the system remains in athermal mechanical equilibrium (i.e. along the athermal
elastic branch) the force $\B f_i$ on every particle is zero before and after an infinitesimal deformation;
In other words~\cite{10KLLP} with $U$ the potential energy
\begin{eqnarray}
&&\frac{d\B f_i}{d\gamma}=\frac{d}{d\gamma}\frac{\partial U}{\partial \B r_i}=
\frac{d}{d\gamma}\frac{\partial U}{\partial \B u_i}=\\
&&= \frac{\partial^2 U}{\partial\gamma\partial \B u_i}+\frac{\partial^2 U}{\partial\B u_i\partial \B u_j}
\frac{d \B u_j}{d\gamma}
\equiv \B \Xi_i +\B H_{ij}\frac{d \B u_j}{d\gamma}=0 \ , \nonumber
\end{eqnarray}
where summation is implied by repeated indices. 
This condition introduces the all-important Hessian matrix $\B H_{ij}$ and the `non-affine force' $\B \Xi_i$ which
can both be computed from the interparticle interactions. We rewrite this condition as
\begin{equation}
\frac{d\B u_i}{d\gamma}=-\B H_{ij}^{-1} \B \Xi_j = -\sum_k\frac{ \B \psi^{(k)}_j\cdot \B \Xi_j }{\lambda_k}\B \psi^{(k)}_i\approx -\frac{ \B \psi^{(P)}_j\cdot \B\Xi_j }{\lambda_P}\B \psi^{(P)}_i \ , \label{duidt}
\end{equation}
where the second equation results from expanding in the eigenfunctions of $\B H$, $\B H_{ij} \B \psi_j^{(k)} =\lambda_k
\B \psi_i^{(k)}$, and the last estimate stems from our knowledge that in finite systems the plastic event is associated with a single
eigenvalue going through zero when the systems slides over a saddle. We denote the critical eigenvalue as $\lambda_P$.
Eq. (\ref{duidt}) can be integrated to provide the distance of the non-affine field $\B u_i$ from its value at $\gamma_P$,
$
\B u_i(\gamma) - \B u_i(\gamma_P) = X(\gamma) \B \psi^{(P)}_i $,
where $X(\gamma)$ is a function of $\gamma$ only, satisfying
\begin{equation}
\frac{dX(\gamma)}{d\gamma} \approx -\frac{ \B \psi^{(P)}_j\cdot \B\Xi_j }{\lambda_P}\B   \ . \label{dxdg}
\end{equation}
Finally, we use the crucial assumption~\cite{10KLLP} that the eigenvalue $\lambda_P$ crosses zero with a finite slope in the $X$-coordinate system itself, where distances are measured along the unstable direction:
\begin{equation}
\lambda_P \approx A X +{\C O}(X^2) \ , \label{lamX}
\end{equation}
Together with Eq.~(\ref{dxdg}) and asserting that $\B \Xi_j$ is not singular (it is a combination of derivatives of the potential function~\cite{10KLP}), implies that
\begin{equation}
X(\gamma)\propto \sqrt{\gamma_P-\gamma} \ . \label{Xgam}
\end{equation}

These results are now used to determine the singularity of the stress at $\gamma_P$. We start with the exact result
for the shear modulus~\cite{MaloneyLemaitre2004a,LemaitreMaloney2006}
\begin{equation}
\mu \equiv \frac{d\sigma_{xy}}{d\gamma}=\mu_B -\frac{1}{V}\B\Xi \cdot{\B H}^{-1}\cdot \B\Xi \ ,
\end{equation}
where $\mu_B$ is the Born term $\mu_B=V^{-1}\partial^2 U/\partial \gamma^2$. Using Eqs. (\ref{lamX}) and (\ref{Xgam}) we conclude that near $\gamma_P$ we can write
the shear modulus as a sum of a regular and a singular term,
\begin{equation}
\mu \approx \tilde \mu - \frac{a/2}{ \sqrt{\gamma_P-\gamma}} + {\C O} ( \sqrt{\gamma_P-\gamma}) \ .
\end{equation}
Obviously, our second derivative $B_2\equiv \frac{d^2\sigma_{xy}}{d\gamma^2}$ will inherit the singularity
from the first derivative, explaining the long tails of the distributions seen in Fig. \ref{pdf}. Close to
mechanical instabilities $B_2$ is expected to diverge like $(\gamma_P-\gamma)^{-3/2}$ with a sign that depends
on the direction of imposed strain.

Even though we sample our configurations from system with temperature where the singular points are not reached
due to thermal activations, the proximity of these mechanical instabilities in a specific straining direction is signalled by the large values of $B_2$ that we measure.

\section{summary and conclusions}

We have proposed here a new measure of the deformation-history induced anisotropy in amorphous solids. This measure is not
model dependent and is easily accessible to simulations and experiments. 
It is not obvious at this point in time whether a theory of elasto-plasticity should take into account the full pdf
of $B_2$, or whether it
would be sufficient to take the mean value of $B_2$ into account. We propose however that this object and its
pdf are tempting analogues of the object $m$ of the STZ theory as discussed above, with the obvious advantage
that they can be easily measured. In fact, in a follow up paper \cite{10KLP} we will show that this object can be expressed as a sum over the particles in the system, and therefore the measurements of the pdf can be done naturally and rapidly,
making them highly accessible for further research. We stress that the value of $B_2$ which has been defined as
the limit $T\to 0$ in Eq. (\ref{defB2}) can be measured experimentally at sufficiently low temperatures where the Bauschinger effect is expected to be saturated. It appears worthwhile to measure this quantity in such low-temperature
experiments and to correlate the value with the amplitude of the Bauschinger effect.

\acknowledgments
We thank Noa Lahav for her help with the graphics, and to Eran Bouchbinder and Anael Lema$\^{\mbox i}$tre for useful discussions.
This work has been supported by the Israel Science Foundation and the German Israeli Foundation.

\end{document}